\input phyzzx\def\e{\adveq\eqno{\rm (\chapterlabel\the\equanumber)}}

\def\adveq{\global\advance\equanumber by 1}
\def\myeq{{\rm \chapterlabel\the\equanumber}}
\def\rarrow{\rightarrow}

\def\semidirect{\mathrel{\raise0.04cm\hbox{${\scriptscriptstyle |\!}$
\hskip-0.175cm}\times}}

\def\mod{\mathop{\rm mod}\nolimits}

\def\ref#1{$^{[#1]}$}

\def\r#1{$[\rm#1]$}

\def\e{\adveq\eqno{\rm (\chapterlabel\the\equanumber)}}

\def\adveq{\global\advance\equanumber by 1}
\def\myeq{{\rm \chapterlabel\the\equanumber}}
\def\rarrow{\rightarrow}

\def\semidirect{\mathrel{\raise0.04cm\hbox{${\scriptscriptstyle |\!}$
\hskip-0.175cm}\times}}

\def\mod{\mathop{\rm mod}\nolimits}

\def\ref#1{$^{[#1]}$}

\def\r#1{$[\rm#1]$}

\date{June, 2006}
\date{June 2006}
\titlepage
\title{Galois Groups in Rational Conformal Field Theory}
\author{Doron Gepner}
\line{\hfill Department of Particle Physics\hfill}
\line{\hfill Weizmann Institute\hfill}
\line{\hfill Rehovot 76100, Israel\hfill}
\abstract
It was established before that fusion rings in a rational conformal field theory (RCFT) can be
described as rings of polynomials, with integer coefficients, modulo some relations.
We use the Galois group of these relations to obtain a local set of equation for the points of
the fusion variety. These equations are sufficient to classify all the RCFT, Galois group by Galois group.
It is shown that the Galois group is equivalent to the pseudo RCFT group. We prove that the
Galois groups encountered in RCFT are all abelian, implying solvability by radicals of the modular matrix.
\endpage
A great deal of interest in conformal field theory stems from the complete solvability of the theories, 
see examples of such theories in  \REF\RCFT{A.A. Belavin, A.M. Polyakov and A.B. Zamolodchikov, 
Nucl. Phys. B 241 (1984) 33;
V. Knizhnik and A.B. Zamolodchikov, Nucl. Phys. B 247 (1984) 83.}
\REF\GW{D. Gepner and E. Witten, Nucl. Phys. B 278 (1986) 493.} refs. \r{\RCFT,\GW}.

Conformal field theories in two dimensions arise in phase transitions as well as string theory. Here we address the
problem of classifying such theories, using Galois groups. This would help in the classification of the
possible universality classes in two dimensional critical phenomena, as well as possible string theories.

An important notion in rational conformal field theory (RCFT) is the fusion ring. Assume that $[a]$ and $[b]$ are some
primary fields. Then in the operator product expansion (OPE) of these two fields we find several conformal blocks,
labeled by the primary field $[c]$,
$$[a](z) [b](w)=\sum_c (z-w)^{\Delta_c-\Delta_a-\Delta_b} [c](w)+\ldots,\e$$
where $\Delta_a$ is the dimension of the primary field $[a]$.
Thus, we define the fusion ring according to the fields appearing in this OPE,
$$[a] \times [b]=\sum_c N_{ab}^c [c],\e$$
where the coefficients $N_{ab}^c$ are non negative integers, obeying commutativity and associativity. We conclude 
that the fusion rules define a commutative ring, denoted by $R$.

For an example, take the $SU(2)_k$ WZW theory. Here the primary fields are labeled by twice the isospin,
$[j]$ such that $j=0,1,2,\ldots,k$. The fusion ring is given by the ``depth" rule, ref. 
\r\GW,
$$[j]\times [m]=\sum_{l\geq |j-m| \atop j+m+l=0\mod2}^{\min(j+m,2 k-j-m)} [l].\e$$
It follows that this ring is generated by $x=[1]$ and we have, from the fusion rules, eq. (3), 
$[2]=x^2-1$, $[3]=[1] [2]-[1]=x^3-2 x$, etc. The general formula for these polynomials are the Chebishev polynomials 
of the second kind,
$$[m]=p_m(x) \qquad {\rm obeys\ } p_m(2 cos \phi)={\sin (m+1)\phi \over \sin \phi},\e$$
where $\phi$ is some variable. 

Now, the relation for this ring follows from the equation $p_{k+1}(x)=0$.
We conclude that in this case the fusion ring can be presented as the quotient ring
$$R\approx {Z[x] \over (p_{k+1}(x))},\e$$
where $Z[x]$ are the polynomials with coefficients in $Z$ (the integers) and $(p_{k+1}(x))$ denotes the 
ideal generated by the relation $p_{k+1}(x)$ (i.e., all the products of this polynomial by any other polynomial in the
ring).

To demonstrate, take $k=2$. We then have the primary fields $[0]=1$, $[1]=x$ and $[2]=x^2-1$, along with
the relation $p_3(x)=x^3-2 x$.

Consider some general fusion ring, $R$. As we shall prove it is semi--simple, i.e., the intersection of all the
maximal ideals, which is called the Jacobson radical, and denoted by $J$, is trivial,
$$J=(0).\e$$

Another radical that we may define in a commutative rings is the Nil radical which is the ideal comprised of all 
the nilpotent elements in the ring, i.e., elements $x\in R$ such that $x^m=0$ for some integer $m$. The Nil
radical is always contained in the Jacobson radical (for an explanation and references see 
\REF\DGSB{D. Gepner, Commun. Math. Phys. 
141 (1991) 381}\r\DGSB). In the case of finitely generated commutative rings the vanishing of the Nil radical
is equivalent to the vanishing of the Jacobson radical. So, we shall refer to this ring as semi--simple.  
For a commutative semi--simple ring this implies that there are no nilpotent elements.

Now, there is a standard map from any commutative ring to the ring of polynomials. Suppose that $p_1$, $p_2$,...,$p_n$
are some generators of this ring (we have a finite number of generators since this 
is a finitely generated ring). We than have the onto ring
homomorphism $h$,
$$h: Z[x_1,x_2,\ldots,x_n]\rarrow R,\e$$
where any polynomial is sent to the element of $R$ defined by substituting the generator $p_r$ for $x_r$. Denote by
$I$ the kernel of this map, $I={\rm ker}(h)$. This is the ideal of all polynomial relations, i.e., the polynomials
vanishing in the ring $R$. According to the first isomorphism theorem, the ring $R$ is isomorphic to this quotient ring,
$$R\approx {Z[x_1,x_2,\ldots, x_n]\over I}.\e$$

Denote by $V$ all the points for which all the polynomials in $I$ vanish,
$$V=\left\{ (x_1,x_2,\ldots,x_n )\,\bigg|\, p(x_1,x_2,\ldots,x_n)=0\quad{\rm for \ all \ } p\in I\right \}.\e$$
This collection of points (of finite number) which are in the algebraic closure of $Z$ (which we conveniently take to be
the complex numbers $C$), we call the fusion variety.

Hilbert's Nullstelensatz theorem asserts that any polynomial vanishing on the points of $V$, $p(V)=0$, obeys
$p^m\in I$ for some positive integer power $m$. In our case, since there are no nilpotent elements, this implies that
$I$ can be described as the ideal of all polynomials vanishing on the fusion variety $V$,
$$I=\left\{ p\in Z[x_1,x_2,\ldots,x_n]\, \bigg |\, p(V)=0\right\}.\e$$

The ring $R$ can be thought of as the ring of polynomial maps, with integer coefficients, from the fusion variety,
$V$, to the complex numbars, $C$. The map is obtained by evaluating any polynomial element of the ring, $p$, 
on the points of the variety, 
$$p\in R\longrightarrow p(x^\alpha)\in C, \quad {\rm where\ } x^\alpha\in V,\e$$
and we used a vector notation for the points of the variety, $x^\alpha=(x_1^\alpha,x_2^\alpha,\ldots, x_m^\alpha)$.
This is a well defined map on the ring $R$, since precisely on these points the ideal $I$ vanishes, and so
the answer does not depend on the quotient by $I$. This is, in fact, a category map, from the category of commutative,
semi--simple rings, to the category of polynomial maps on some variety, with integer coefficients. 
Homomorphisms of the ring correspond
to the composition of the maps.

We can make use of Verlinde formula, ref. \REF\Verlinde{E.P. Verlinde, Nucl. Phys. B 300 (1988) 360.}\r\Verlinde,
at this stage. Define the polynomial field in the ring $R$ by 
$$\beta_i=\sum_i S_{ir} [r],\e$$
where $S$ is the modular matrix. According to Verlinde equation, we have the product of polynomials as follows,
$$\beta_i \beta_j=\delta_{ij} \lambda_i \beta_i,\qquad {\rm where\ } \lambda_i=S_{i0}^{-1}.\e$$
Substituting any of the points of the variety $V$, we obtain a product of complex numbers,
$$\beta_i(x^\alpha) \beta_j(x^\alpha)=\lambda_i \delta_{ij} \beta_j(x^\alpha).\e$$
This implies that each of the $\beta_i(x^\alpha)$ is $0$ on all the points of the variety, except for one point,
where it is $1$. Defining a basis of $R$ by the polynomial map $\zeta_i$, which obeys,
$$\zeta_i(x^j)=\delta_{ij},\e$$
we see that we may write the polynomials $\beta_i$ as,
$$\beta_i =\zeta_i \lambda_i.\e$$
In particular, this formula, eq. (16), implies that the fusion ring, $R$, is semi--simple, since we have 
exactly $n$ independent polynomials for a ring of dimension $n$, and so the variety of points vanishing in $I$, 
cannot be degenerate.

Using the unitarity of $S$, we may invert the relation, eq. (12), and we find the value of any primary field
on the points of the fusion variety,
$$p_r(x_i)={S_{ri}^\dagger \over S_{0i}},\e$$
where the $0$ denotes the unit primary field, $[0]$. Now, since the generators, denoted by $1, 2,\ldots,m$,
are some primary fields, we have for these the relation, $x^a=p_a(x)$, and from eq. (15),
$$x^a_i={S^\dagger_{a i} \over S_{0i}},\e$$
and we proved the theorem, ref. \r\DGSB,

Theorem (1): 
Any fusion ring is the ring of polynomials $Z[x_1,x_2,\ldots,x_m]/I$ where $I$ is the ideal of all polynomials 
vanishing on the fusion variety $x^i_\alpha={S_{i\alpha}^\dagger\over S_{i0}}$,
where $\alpha=1,2,,\ldots,m$ labels the generators, and $i$ denotes the different points of the variety,
whose number is the number of primary fields.
Furthermore, the value of any
primary field $[a]$ evaluated on any of the points of the fusion variety is given by, 
$p_a(x_i^1,x_i^2,\ldots,x_i^m)={S_{a,i}^\dagger\over S_{0,i}}$. 

Let us return to our example of $SU(2)_k$. Here we have only one generator, $x=[1]$. For simplicity,
assume that $k=3$. The relation in this case gives rise to the equation,
$x^3-2 x=0$. The fusion variety $V$ corresponds here to solutions
of this equation, which are, $x_0=\sqrt2,\ x_1=0, x_2=-\sqrt2$. The ideal of relations can then be described as all the 
polynomials vanishing on these three points.

In the case where there is only one generator, $m=1$, we have a ring of polynomials with one variable, $Z[x]$. We
know that this ring is a principle ideal domain (PID), which means that any ideal is generated by some one polynomial,
$q(x)$. And so any such ring is the quotient,
$$R\approx {Z[x]\over (q(x))}.\e$$

From hereon we will mainly concentrate on the one generator case. In this case, we can assume that the primary field
$p_m(x)$ is a polynomial of degree $m$ in $x$ where the leading coefficient is $1$, $p_m(x)=x^m+a_{m-1} x^{m-1}+\ldots$,
where the coefficients $a_i$ are integers. For simplicity, we shall also assume that the ring is real, i.e., the modular
matrix $S$ is real,
$$S=S^*.\e$$

Denote the points of the fusion variety by $x_0,x_1,\ldots,x_{n-1}$. The second primary field, $p_2$ is then given by 
$$p_2(x)=x^2-1-b x,\e$$
since, in the fusion ring, $[1]\times [1]=[2]+[0]+b [1]$ (the [0] must appear since this field is self conjugate),
and where $b$ is any non--negative integer. In all examples we can see that $b=0$ or $b=1$, so we shall adhere
mainly to these cases. 

We can define the matrix by $M_{i\alpha}=S_{i\alpha}^\dagger/S_{i0}$ and from theorem (1), and it is given by,
$$M_{i\alpha}={S_{i\alpha}^\dagger \over S_{i0}}=\pmatrix{1 & 1 &\ldots &1\cr x_0 & x_1 &\ldots &x_{n-1} \cr
p_2(x_0) & p_2 (x_1) &\ldots & p_2(x_{n-1}) \cr \vdots & \vdots &\ddots & \vdots\cr p_{n-1}(x_0) & p_{n-1}(x_1) 
&\ldots &p_{n-1}(x_{n-1}) \cr}.\e$$

In an RCFT, the modular matrix is a uniatry and symmetric matrix. The symmetry of the $S$ matrix,
$$S_{ij}=S_{ji},\e$$
simple as it may seem, is, in fact, a very powerful constraint. We may define as a symmetric affine
variety, any variety (or collection of points) for which the points of the variety obey eq. (22) with
a symmetric unitary matrix. This condition alone is enough, it seems, to classify completely all the fusion rings.
Or, we conjecture that every symmetric affine variety is the fusion variety of some full fledged RCFT.

Define, $p_0(x)=1$, $p_1(x)=x$, and assume that
$$\lambda_r^{-2}=\sum_{m=0}^{n-1} | p_m(x_r) |^2.\e$$
Then from the unitarity and the symmetry of the modular matrix, $S$, we get the following very important set of 
equations on the fusion variety,
$$\lambda_r p_s(x_r)=\lambda_s p_r(x_s),\e$$
for all $r,s=0,1,\ldots,n-1$. This relation, which is very constraining, is the basis for our classification
of symmetric affine varieties, or, fusion rings.

Taking the cases where $(r,s)=(0,1), (0,2), (1,2)$, we get the following three equations,
$$\eqalign{x_0\lambda_0&=\lambda_1, \cr \lambda_0 p_2(x_0)&=\lambda_2,\cr \lambda_1 p_2(x_1)&=x_2 \lambda_2.\cr}\e$$
Where, recalling from before, $p_2(x)=x^2-b x-1$, where $b=0$ or $b=1$.

Combining these three equations, we arrive at the relation,
$$x_0 p_2(x_1)=x_2 p_2(x_0),\e$$
which is a special case of
$$p_r(x_0) p_s(x_r)=p_s(x_0) p_r(x_s),\e$$
for any $r,s\geq 1$.
The importance of these relations is that they are local, i.e., they do not depend on the entire modular matrix, but
only on these particular three points, where $p_2$ is already known to us explicitly.

As before, we define the relation for this ring by the polynomial $q(x)$. We recall now some elements of Galois theory,
as applied to this polynomial. Denote, as before, the roots of this polynomial by $x_0,x_1,\ldots,x_{n-1}$.
We define the Galois field of this polynomial by the field generated over the rational numbers, $Q$, by the
roots $x_0,x_1,\ldots,x_{n-1}$,
$${\cal F}=\left\{ \sum_i q_i x_i \,\bigg |\,  q_i {\rm\  rational} \right \}.\e$$
The field, $\cal F$ is called the Galois field of the equation $q(x)=0$.

Denote by $G$ the group of all the automorphisms of the Galois field $\cal F$. An automorphisms is a map,
$\sigma$, from
$\cal F$ to itself,
$$\sigma: {\cal F}\rightarrow {\cal F},\e$$
which is one to one and onto and preserves the product and the addition in the field,
$$\eqalign{\sigma(x y)=&\sigma(x) \sigma(y),\cr \sigma(x+y)=&\sigma(x)+\sigma(y).\cr}\e$$
The automorphisms form a group under decompositions. This group is called the Galois group of the equation.

Now, acting on the equation $q(x)=0$ with an automorphism $\sigma$ we get $q(\sigma(x))=0$. Since $x_0,x_1,\ldots,x_{n-1}$
are the roots of this equation, it follows that the Galois automorphism permutes the roots,
$$\sigma(x_r)=x_{p(r)},\e$$
where $p$ is some permutation. It follows that the Galois group is a subgroup of the permutation group $S_n$, the
permutations of $n$ objects. The `typical' equation will have the full $S_n$ as its Galois group.

A well known theorem by Galois asserts that if the Galois group is solvable then the equation may be solved by radicals
(i.e. some roots). The solvability of the group is defined by a chain of normal subgroups,
$$(1)\subset G_m\subset G_{m-1}\subset\ldots\subset G_1\subset G,\e$$
(where $G_0=G$, $G_{m+1}=(1)$ the trivial group),
such that the quotients $G_i/G_{i+1}$, where $i=0,1,2.\ldots,m$, are all abelian groups. 
In particular, any equation with an abelian Galois group
is solvable by radicals.

Let us consider the Galois group of the fusion variety, with the equation $q(x)=0$. Acting on eq. (25) with
any automorphism $\sigma\in G$, $\sigma(x_r)=x_{a(r)}$, where $a$ is some permutation, we find
$$\lambda_{a(r)} p_{m}(x_{a(r)})=\lambda_{a(m)} p_r(x_{a(m)} ),\e$$
where we have the freedom of flipping the sign of $\lambda_{a(r)}$ owing to the square root in its definition, 
eq. (24).
In particular, eq. (27) becomes,
$$x_{a(0)} (x^2_{a(1)}-1-b x_{a(1)} )=x_{a(2)} (x^2_{a(0)}-b x_{a(0)}-1 ),\e$$
where the coefficient $b$ is zero or one. This equation holds for any permutation $a$ in the Galois group,
$a\in G$.
Thus, instead of one single equation, eq. (27), we have a system of equations whose number is the order of the Galois
group. This is of great importance to us, since, as we shall see, it is exactly enough equations, in any physical
situation, to exactly solve for the points of the variety. So, this set of equations would enable
us to classify fusion rings.

To exemplify, let us return to the case of $SU(2)_3$. In this RCFT we have three elements in the fusion variety,
$x_0$, $x_1$ and $x_2$, along with a $Z_2$ Galois group, $\sigma(x_0)=x_2$, $\sigma(x_2)=x_0$.
So let us assume this Galois group, and also that $p_2(x)=x^2-1$, i.e.,
$b=0$, and no further assumptions.

We than get the equations, from eq. (35),
$$\eqalign{x_0(x_1^2-1)=& x_2 (x_0^2-1),\cr x_2 (x_1^2-1)=& x_0 (x_2^2-1).\cr}\e$$
So,
$$x_1^2-1={x_2 (x_0^2-1)\over x_0}={x_0 (x_2^2-1)\over x_2}.\e$$
We find that $x_0^2=x_2^2=\alpha^2$ or that $x_0=\alpha$ and $x_2=-\alpha$ (since the points of the variety all
have to be distinct).
Now the symmetric polynomial
$$x_0 x_1+x_0 x_2+x_1 x_2,\e$$
is a coefficient in the polynomial relation $q(x)$, and so it has to be an integer. Since its value is $-\alpha^2$,
we infer that $\alpha^2$ is an integer.
From, eq. (36) we have
$$x_1=\sqrt{2-\alpha^2},\e$$
so the only possible real values for $x_1$ are obtained when $\alpha^2=0,1$ or $2$. However, 
$\alpha^2=0$ or $1$ give a degenerate variety, and
thus the only allowed value is $\alpha^2=2$. To summarize, we found the points of the variety and they are given by
$$x_0=\sqrt2,\qquad x_1=0,\qquad x_2=-\sqrt2,\e$$
which are precisely the points of the fusion variety of $SU(2)_2$, as we found before. 
The conclusion is
that we classified completely real fusion rings with a third order polynomial relation, a $Z_2$ Galois group, and the 
relation $[2]=[1]^2-[0]$, and we proved that the only possibility for such a ring is $SU(2)_2$.

This example shows that, indeed, eq. (35), is indeed enough to find completely all the candidates for an RCFT.
Remarkably, it is always the same equation with only the indices of the points changing. We are looking for solutions
to these equations, $x_0,x_1,\ldots,x_{n-1}$ which are the points of the fusion variety. Since these are the
roots of the relation $q(x)=0$, which is an integer coefficient polynomial, the symmetric polynomials of the $x_i$ all 
have to be integers.  The symmetric polynomials are defined by
$$W_r=\sum_{i_1<i_2<\ldots<i_r} x_{i_1} x_{i_2} \ldots x_{i_r},\e$$
and have to be integers. 

Thus, we are classifying RCFT, Galois group by Galois group. It is very easy to solve eq. (35) by Mathematica,
requiring that all $W_r$ are integers. 
In the appendix a sample program is described which solves the eqs. (35), for the theory $SU(2)_9/Z_2$,
proving that it is the only real theory with five primary fields, $b=1$ and a $Z_5$ Galois group.

Recall the notion of `pseudo' conformal field theory, refs. \REF\Pseudoone{D. Gepner, Caltech preprint CALT-68-1825, Hep-th 9211100
(1992).} \REF\Pseudo{E. Baver, D. Gepner and U. Gursoy, Nucl. Phys. B 557 (1999) 505; D. Gepner, Nucl. Phys. B 561 (1999) 467;
E. Baver, D. Gepner and U. Gursoy, Nucl. Phys. B 561 (1999) 473.}\r{\Pseudoone,\Pseudo}.  
Take any conformal field theory, which has the 
modular matrix, $S$, which is a symmetric and unitary matrix. A pseudo conformal field theory is defined by
some permutation $p$, such that the new modular matrix is,
$$S^{\rm new} _{i,j}=S_{i,p(j)} s_j,\e$$
where $s_j$ is a sign dependent only on $j$, $s_j=\pm1$.
This is the modular matrix of some other conformal field theory, the `pseudo' theory. Of course, $S^{\rm new}$
has to be, as well, a symmetric and unitary matrix (unitarity is automatic). The pseudo conformal field
theory has the same fusion rules as the `old' theory. This is a consequence of the Verlinde formula, ref. \r\Verlinde, 
which states that the fusion coefficients are given by,
$$N_{ij}^k=\sum_{\alpha} {S_{i\alpha} S_{j\alpha} S_{k\alpha}^\dagger \over S_{0\alpha}},\e$$
where $S$ is either the old or new modular matrix. Now, since $S^{\rm new}$ is just a permutation of the columns of
$S$ (up to the sign in eq. (42)), the same fusion rules are obtained in eq. (43) for both theories.

For an example of pseudo conformal field theory, consider again the WZW model $SU(2)_k$. The modular matrix is
given by, ref. \r\GW,
$$S_{l,s}=\sqrt{2\over k+2} \sin \left[{\pi (l+1) (s+1)\over k+2}\right ],\e$$
where $l$ and $s$ are twice the isospin of the primary fields. 
Take now the `new' modular matrix,
$$S_{l,s}^{\rm new}=\sqrt{2\over k+2} \sin\left[ {\pi r (l+1) (s+1) \over k+2}\right],\e$$
where $r$ is any integer starange to $2(k+2)$, $\gcd(r,2(k+2))=1$.

Clearly, the new modular matrix is a symmetric matrix. Further, the multiplication by $r$ is equivalent to 
some permutation of the rows, $p$,
$$\sin\left[ {\pi r (l+1) (s+1)\over k+2} \right]=t_s \sin \left[ {\pi (l+1) (p(s)+1)\over k+2 }\right],\e$$
where $t_s=\pm1$ is a sign dependent only on $s$. We may write for the permutation,
$$p(s)+1=\pm r (s+1) \mod 2(k+2).\e$$
So, $S^{\rm new}$, is an example of a pseudo conformal field theory, and, in particular, it has the
same fusion rules as the usual $SU(2)$. The evidence, so far, indicates that any `pseudo' conformal field theory
can indeed be realized by some full fledged RCFT.

The pseudo conformal field theories evidently form a group under the decomposition of the permutations used
in their definitions. In the case of $SU(2)_k$ discussed above, this is simply the group of multiplication
of all integers, $r$, strange to $2(k+2)$ modulo $2(k+2)$, where we identify $r$ and $-r$ (since $-r$ gives the
same pseudo conformal field theory as $r$). In particular, this group is abelian.

Suppose that $\sigma$ is any automorphism in the Galois group. By acing with $\sigma$ on eq. (25) we 
can define,
$$S_{l,m}^{\rm new}=S_{l, p(m)} s_m,\e$$
where $s_m$ is a sign, $s_m=\pm1$,
and $p(m)$ is the permutation of the automorphism, $\sigma (x_m)=x_{p(m)}$. Evidently, $S^{\rm new}$ is a symmetric
matrix, by acting with $\sigma$ on eq. (28), and so it defines a pseudo conformal field theory. It follows that any element
of the Galois group gives rise to a pseudo RCFT of the original one. The converse of this statement is obviously
also correct and any pseudo RCFT, since it preserves the symmetry of the modular matrix, eq. (34) must hold,
and it follows that it corresponds to some Galois automorphism. Thus, we have established the following.

Theorem (2):
The Galois group of a fusion variety is identical to the group of pseudo conformal field theories
based on the particular RCFT.

Evidently, the Galois group is independent of the generator chosen, and it is always the exact same permutation
on the points of the fusion variety.

This allows for an easy calculation of the Galois group in any known RCFT, since their pseudo RCFT are already known,
refs. \r{\Pseudoone,\Pseudo}.
Also, this enables an easy generalization of the notion of a Galois group to more than one variable. We noted above
that for $SU(2)_k$, the Galois group was abelian. It can be seen, further, that in all the known examples of RCFT, the
Galois group is also abelian. This leads to,

Theorem (3):
The Galois group of any RCFT is always abelian.

proof: We know that $S^{\rm new}=S_{l,p(m)}$ is symmetric, where $p$ is some permuatation. We ignore the signs
in eq. (42), as they are irrelvant for this discussion). This implies that $S_{l,p(m)}=S_{m,p(l)}=S_{p(l),m}$.
Now, combining two such permuatation $p$ and $q$, we obtain, $S_{l,pq(m)}=S_{p(l),q(m)}=S_{qp(l),m}=S_{l,qp(m)}$.
We conclude that $pq=qp$, since the rows of $S$ are all different, proving the the pseudo group, 
or the Galois group, which is the same, of any RCFT is abelian, for any number of generators.

In particular, since the Galois group is solvable (actually, abelian), this implies, by Galois theory, 
that the relations
encountered in an RCFT are always solvable by radicals. In particular, the elements of the modular matrix can always
be expressed by roots. This is quite a fascinating observation, especially, since, it is not always easy
to see how this is done in practice.

It is actually not difficult to see that the Galois group cannot be the `typical' permutation group $S_n$ (for $n>3$).
In this case, eq. (35) entails that
$$x_{p(0)} (x_{p(1)}^2-1-b x_{p(1)})=x_{p(2)} (x_{p(0)}^2-1-b x_{p(0)}),\e$$
for any permutation $p\in S_n$. Clearly, we have many more equatios than unknowns, and so we cannot
expect to have solutions, which are all distinct. Assume that the number of points, $n$, is greater than $3$. 
Suppose also that the permutation is $(2,s)$ where $s$ obeys $s>2$. Then eq. (49) would imply immediately that
$x_2=x_s$, by substituting this permutation. 
($p_2(x_0)$ which is $\lambda_2/\lambda_0$ cannot be zero because of eq. (24).)
Since our ring is semi--simple, it is not possible to have degenerate
values for the variety. Similarly, it can be seen directly that for the Galois group $S_3$ (where $n=3$) there
are no solutions (which are not degenerate) and we proved,

Statement: The Galois group in an RCFT cannot be the `typical' full permutation group for $n\geq3$ elements.

Actually, we proved the stronger statement that the Galois group cannot contain the permutation $(2,n)$.

All of these considerations can quite readily be generalized to more than one generator. E.g., suppose
that there are two generators denoted by $x$ and $y$ (which are some primary fields).
Then, we have for the matrix $M$,
$$M_{i a}={S_{ia}^\dagger\over S_{i0} }=\pmatrix{1&1&\ldots&1 \cr x_0 & x_1 &\ldots & x_{n-1} \cr y_0 &y_1 
&\ldots &y_{n-1}\cr p_2(x_0,y_0) & p_2(x_1,y_1) &\ldots & p_2 (x_{n-1}, y_{n-1} )\cr
\ldots & \ldots &\ldots &\ldots \cr},\e$$
where, $p_2(x,y)=x y -n_1 x-n_2 y$ and $n_1$ and $n_2$ are some fixed non--negative integers. Again, the symmetry
of the $S$ matrix gives us enough equations, the analogue of eq. (34), 
when combined with the Galois group, to classify
the possible theories. The notion of the Galois group is easily generalized to more than one variable
by taking the equivalent pseudo CFT group.

Indeed, eq. (28) generalizes easily to more that one generator, by simply taking $x_r$ as a vector,
$x_r=(x_r^1,x_r^2,\ldots,x_r^m)$, where there are $m$ generators. And so,
$$p_r(x_0^1,x_0^2,\ldots,x_0^m) p_s(x_r^1,x_r^2,\ldots,x_r^m)=p_s(x_0^1,x_0^2,\ldots,x_0^m) p_r(x_s^1,x_s^2,\ldots,x_s^m).\e$$
The pseudo conformal group permutes the columns of the $S$ matrix. So, it corresponds to a permutation,
$r\rarrow a(r)$ where $r$ is any of the points of the variety, and we take, $x_r^b\rarrow x_{a(r)}^b$, where,
very importantly, it is the same permuatation for all the generators. Thus, we have again, a system of equations,
by applying any permutation, $a$, in the pseudo conformal group,
$$p_r(x_{a(0)}) p_s(x_{a(r)})=p_s(x_{a(0)}) p_r(x_{a(s)}),\e$$
where we used again the vector notation.

To exemplify, for two fields, we have the equations, substituting the modular matrix, eq. (50),
$$\eqalign{y_{a(1)} x_{a(0)}&=x_{a(2)} y_{a(0)},\cr
x_{a(0)} p_2(x_{a(1)},y_{a(1)})&=x_{a(3)} p_2(x_{a(0)},y_{a(0)}),\cr
y_{a(0)} p_2(x_{a(2)},y_{a(2)})&=y_{a(3)} p_2(x_{a(0)},y_{a(0)}),\cr}\e$$
where $a$ is any permutation in the pseudo conformal group. We can solve these equations to get the variety
points, $x_r$ and $y_r$, since we know $p_2(x,y)$.

We believe that the fact that the pseudo conformal group, or the `Galois' group is abelian, for more than
one generator, also implies the solvability by radicals of the fusion variety, or the modular matrix.
However, this remains to be proved.

Let us return to the one generator case. Once we found a solution to eq. (35), and we have the fusion variety,
we need to construct the full modular matrix. This we do now by proceeding in a different way. We know that 
$$p_3(x)=x p_2(x)-b_1 x-b_2 p_2(x),\e$$
where $b_1$ and $b_2$ are some non--negative integers. From eq. (25) we have,
$$\lambda_r p_s(x_r)=\lambda_s p_r(x_s).\e$$
Taking the values, $r=3$ and $s=2$, and $r=3$ and $s=1$, we find the equations,
$$\eqalign{\lambda_2 p_3( x_2)=&\lambda_3 p_2 (x_3),\cr
\lambda_1 p_3(x_1)=&x_3 \lambda_3,\cr}\e$$
and substituting the values of the $\lambda$'s, eq. (25), we get the relations,
$$\eqalign{p_2 (x_0) p_3 (x_2)&=p_3(x_0) p_2 (x_3)\cr
x_0 p_3(x_1)=&x_3 p_3(x_0),\cr}\e$$
which is a specialization of eq. (28).
There are exactly two equation for two unknowns, i.e., the coefficients $b_1,b_2$, in eq. (54). 
(Actually, it is a linear equation, 
so we have exactly one solution.) If it does not come out a non--negative integer, we discard the theory. 
If it is an integer, we proceed by
taking 
$$p_4(x)= x p_3 (x) -b_1 p_2 (x)-b_2 x-b_3 p_3(x),\e$$
where $b_1$, $b_2$ and $b_3$ are some non--negative integers. Again, we have three equations for three unknowns, 
from eq. (28),
and so we can easily
solve for the coefficients, $b_1$, $b_2$ and $b_3$. 
We proceed in this fashion, until we fully construct
the modular matrix. Then, we check that the matrix is unitary, and that the full fusion coefficients, calculated
by Verlinde formula, eq. (43), are all non negative integers. 
There are, in fact, solutions to eqs. (35) which do not give rise
to a unitary modular matrix, and so must be discarded. (We found one such solution for $n=6$ and $b=1$.)

This way, we classify RCFT Galois group by Galois group. The solution depends only on the particular Galois group
and not on its realization on the fusion variety, since it is the same, up to  a permutation, giving the
same relations in the ring. Also, the Galois group, being equivalent to the pseudo CFT group, is independent
of the particular choice of a generator (or generators).

Actually, for some `small' Galois groups we may not have enough equations to get a unique solution. From examining the
possible theories, it appears that such Galois groups do not arise in RCFT. For example, if $G$ is trivial, the
elements of the $S$ matrix must all be rational, which never happens, for $n>2$. 
It appears that we always have enough equations
to solve uniquely, with a physical Galois group. 
Thus, we can limit ourselves to `sensible'
Galois groups which seem to be the only ones occurring in an RCFT.

For $n\leq 6$ we looked on all Galois groups with one generator. We found that all the solutions are the known ones,
classified in ref. \REF\GK{D. Gepner and A. Kapustin, Phys. Lett. B 349 (1995) 71.}\r\GK,
by a numeric procedure. 
So, we, in fact, proved here rigorously, this tentative classification.
In the table we list all the theories which have one generator, up to six primary fields, along
with their Galois groups. In particular, the Galois groups are all abelian, in accordance with theorem (3).
We see that the order of the Galois group is always less or equal to the number of primary fields. 
This is important in order to have a solution to eqs. (35), and we presume that it is true in general.

We hope that the present work will be a valuable tool in classifying conformal field theories. As
we demonstrated, this classification proceeds Galois group by Galois group. We mainly concentrated
here on the one generator fusion rings. In fact, there is no difficulty in generalizing this classification
to theories with more than one generator, as mentioned above. 
We believe that the results presented here are the first step
in the direction of such a classification.

\overfullrule=0pt

%
%
\setbox\strutbox=\hbox{\vrule height12pt depth8pt width0pt}
\midinsert
\line{\hfill Table. \hfill}
\line{\hfill The Galois groups for one generator theories with up to six primary fields.\hfill}
\vskip15pt
\line{\hfill
\vbox{\offinterlineskip\hrule
\halign{&\vrule#&
  \strut\quad\hfil#\hfil\quad\cr
height2pt&\omit&&\omit&&\omit&\cr
&Number of fields&&Theory&&Galois group&\cr
\noalign{\hrule}
height2pt&\omit&&\omit&&\omit&\cr
&2&&$SU(2)_1$&&$Z_1$&\cr
&2&&$SU(2)_3/Z_2$&&$Z_2$&\cr
&3&&$SU(2)_2$&&$Z_2$&\cr
&3&&$SU(2)_5/Z_2$&&$Z_3$&\cr
&3&&$SU(3)_1$&&$Z_2$&\cr
&4&&$SU(2)_3$&&$Z_2$&\cr
&4&&$SU(2)_7/Z_2$&&$Z_3$&\cr
&4&&$SU(4)_1$&&$Z_2$&\cr
&5&&$SU(2)_9/Z_2$&&$Z_5$&\cr
&5&&$SU(2)_4$&&$Z_2$&\cr
&5&&$SU(5)_1$&&$Z_4$&\cr
&6&&$SU(2)_5$&&$Z_3$&\cr
&6&&$SU(6)_1$ &&$Z_2\times Z_2$&\cr
&6&&$SU(2)_{11}/Z_2$&&$Z_6$&\cr
height2pt&\omit&&\omit&&\omit&\cr
}\hrule}
\hfill}
\setbox\strutbox=\hbox{\vrule height8.5pt depth3.5pt width 0pt}
\vskip20pt\endinsert

\par

\ack
The author thanks S. Cherkis and A. Schwimmer for helpful discussions.

\refout

\Appendix{}

{\bf An example of $SU(2)_9/Z_2$.}
\def\chapterlabel{A}

Following ref. \r\Pseudoone, we shall describe the conformal field theory
$SU(2)_{2 n-1}/Z_2$ where $n$ is the number of primary fields.
This theory can be thought of as the integer isospin representations
of $SU(2)_{2n-1}$ and it forms a consistent conformal field theory.
In ref. \r\Pseudoone\ realizations for these theories are described.

The modular matrix of the theory is given by
$$S_{l,m}=\sqrt{4\over 2 n+1} \sin 
\left[{\pi (2 l-1) (2 m-1)\over 2 n+1}\right],\e$$
where $l$ and $m$ take the values $1,2,\ldots,n-1$, and
$2 l-1$ is twice the isospin plus one. We shall denote
these primary fields by $[2 l-1]$ for $l=1,2,\ldots,n$.

We can describe now the pseudo conformal field theories associated
to this theory. The new modular matrix, $S$, is given by 
$$S^{\rm new}_{l,m}=\sqrt{4\over 2 n+1} \sin\left[{\pi r 
(2 l-1) (2 m-1)\over 2 n+1}\right],\e$$
where $r$ is any integer strange to $2 n+1$. It follows that the psudo
CFT group is isomorphic to the group of integers strange to
$2 n+1$, $\gcd(r,2 n+1)=1$, under multiplication modulo $2 n+1$.
In particular, the pseudo group, which is equivalent to the
Galois group, is abelian.

For example, take $n=5$, the theory with five primary fields.
It is given by the integer isospin representations of the WZW
model $SU(2)_9$. For a generator of the pseudo group, take $r=2$
in eq. (A2). We then obtain the corresponding permutation of
the fields by multiplying by two the odd integers $1,3,5,7,9$
modulo $11$, where $[m]$ is equivalent to $[11-m]$, and $m$
is twice the isospin plus one.
We find that the permutation of the fields, $2 l-1$, corresponding
to this pseudo theory is,
$$p=(1,9,7,3,5),\e$$
i.e., $[1]$ goes to $[9]$, $[9]$ goes to $[7]$, $[7]$ goes to $[3]$, etc. The permutation
$p$ is an element of order $5$ which generates the Galois group,
which consequently is the group $Z_5$ (in agreement with the 
table).

The fusion rules of this theory are the same as those of the integer 
isospin representations of $SU(2)_{2 n-1}$, which are given by
eq. (3), now restricted to even $j$ and $m$ (and $k=2n-1$),
without any further changes. Using the notation of twice the
isospin plus one, we find that the fusion ring is generated by
one field. There are, in fact, two possibilities for the generator.

1) $x=[3]$. In this case, the next field obeys,
$$[5]=[3]\times [3]-[3]-[1],\e$$
and so we have $p_2(x)=x^2-x-1$. This corresponds to the case
of $b=1$ in eq. (21). 

2) $x=[2 n-1]$. In this case, we have the fusion rule
$$[2 n-1]\times [2 n-1]=[1]+[3],\e$$
and in this case,
$$[3]=x^2-1,\e$$
which corresponds to $b=0$ in eq. (21).  
(In the second case, we must permute the primary fields so 
$[2 n-1]$ is the first field, $[3]$ is the second, and the
rest remain as is.)

Thus, we see that we can use either possibility for $b$ in classifying
this theory. For definiteness, we shall assume $b=1$, which is possibility (1)
above. From the fusion rules, eq. (3), the relation in the ring is
then seen to be 
$$q(x)=x^5-4 x^4+2 x^3+5 x^2-2 x-1,\e$$
i.e., the ring is given by the quotient
$$R\approx {Z[x]\over (q(x))}.\e$$

Now, consider the equations that the fusion variety satisfy.
Denote the points of the fusion variety by $x_1,x_3,x_5,x_7,x_9$,
in accordance with our notation for the primary fields. These 
points are the solutions to the equation $q(x)=0$.

We know that the fusion variety satisfies eq. (35),
$$x_{a(1)} p_2(x_{a(3)})=x_{a(5)} p_2(x_{a(1)}).\e$$
Substituting
for $a$ any permutation in the Galois group, $1,p,p^2,p^3,p^4$, where
$p$ is given by eq. (A3),
we arrive at the following five equations,
$$\eqalign{ x_1 p_2(x_3)&=x_5 p_2 (x_1),\cr
            x_9 p_2(x_5)&=x_1 p_2 (x_9),\cr
            x_7 p_2(x_1)&=x_9 p_2(x_7),\cr
            x_3 p_2(x_9)&=x_7 p_2(x_3),\cr
            x_5 p_2(x_7)&=x_3 p_2(x_5),\cr}\e$$
where $p_2(x)=x^2-x-1$.
Thus, we have five equations for five unknowns.

Now, we want to find all the solutions to eqs. (A10). This
would classify all the real rings with five primary fields and
with $b=1$ and a Galois group $Z_5$. We look for solutions such that
the symmetric polynomials, $W_r$, defined by eq. (41), are all
integers.
            
We solve the equations, eqs. (A10), using Mathematica. The program 
is listed below.

$$\eqalign{&p2[x\_] := x^2 - x - 1;\cr
&eq1 := x1 p2[x3] - x5  p2 [x1];\cr
&eq2 := x9 p2[x5] - x1 p2 [x9];\cr
&eq3 := x7 p2[x1] - x9 p2[x7];\cr
&eq4 := x3 p2[x9] - x7 p2[x3];\cr
&eq5 := x5 p2[x7] - x3 p2[x5];\cr
&veq := \{eq1, eq2, eq3, eq4, eq5, x1 x3 x5 x7 x9 - 1\};\cr
&ff := NSolve[veq == \{0, 0, 0, 0, 0, 0\}, \{x1, x3, x5, x7, x9\}];\cr 
&w1:=x1+x3+x5+x7+x9;\cr
&w2:=x1 x3 + x1 x5 + x1 x7 + 
    x1 x9 + x3 x5 + x3 x7 + x3 x9 + x5 x7 + x5 x9 + x7 x9;\cr
&\{w1,w2\} /. ff\cr}$$

Here we assumed that $W_5=1$. Then, we pick up all integers entries 
in the output, which are solutions with integer symmetric
polynomials. 
The only such output is $\{4,2\}$ which is given when the $x$'s have the values of
the
fusion variety of $SU(2)_9/Z_2$. We can run the program with other integer
values of $W_5$ (which we can assume to be small), 
and we see that there are no additional solutions.
We conclude that $SU(2)_9/Z_2$ is the only real theory with 
$n=5$, $b=1$ and a $Z_5$ Galois group.
        
\bye